\documentstyle[12pt,aaspp4]{article}
\def\u{{\bf u}}
\def\VS{V\'azquez-Semadeni}
\def\'#1{\ifx#1i{\accent"13\i}\else{\accent"13#1}\fi}

\begin{document}

\title{On the Effects of Projection on Morphology}

\bigskip
\bigskip
\bigskip
\bigskip

\author{B\'arbara Pichardo$^1$, Enrique \VS$^1$, Adriana
Gazol$^{2,}$\footnote[3]{Present address: Instituto de Astronom\'\i a, UNAM,
Apdo. Postal 70-264, M\'exico, D.F., 04510, MEXICO},
Thierry Passot$^2$ and Javier Ballesteros-Paredes$^{1}$ } 

\affil{$^1$Instituto de Astronom\'\i a, UNAM, Apdo. Postal
70-264, M\'exico, D.F., 04510, MEXICO}

\affil{$^2$ CNRS, UMR 6529, Observatoire de la C\^ote d'Azur, B.P.\ 4229,
06304, Nice Cedex 4, FRANCE}

\bigskip

\bigskip
\bigskip
\bigskip
\bigskip
\bigskip

\begin{abstract}

We study the effects of projection of three-dimensional (3D) data onto
the plane of the sky by means of numerical simulations of turbulence
in the interstellar medium including the magnetic field, parameterized
cooling and diffuse and stellar heating, self-gravity and rotation. We
compare the physical-space density and velocity 
distributions with their representation in
position-position-velocity (PPV) space (``channel maps''), noting that
the latter can be interpreted in two ways: either as maps of the column 
density's spatial distribution (at a
given line-of-sight (LOS) velocity), or as maps of the spatial
distribution of a given value of the LOS velocity (weighted by 
density). This
ambivalence appears related to the fact that the spatial and PPV
representations of the data give significantly 
different views. First,
the morphology in the channel maps more closely resembles 
that of the spatial distribution of the LOS velocity component 
than that of the density
field, as measured by pixel-to-pixel correlations between
images. Second, the channel maps contain more small-scale structure
than 3D slices of the density and velocity fields, a fact evident
both in subjective appearance and in the power spectra of the
images. This effect may be due to a pseudo-random sampling
(along the LOS) of the gas contributing to the structure in a channel
map: the positions sampled along the LOS (chosen by their LOS
velocity) may vary significantly from one position in the channel map
to the next.

\end{abstract}

\section{Introduction} \label{intro}

In recent years, numerical simulations of magneto-hy\-dro\-dy\-nam\-ic (MHD)
interstellar turbulence in a variety of regimes have been presented by
several groups. Some of them have modeled the atomic and
ionized ISM at large scales including self-gravity, parameterized heating and
cooling and stellar-like forcing from ionization heating (Passot, \VS\ 
\& Pouquet 1995; hereafter Paper I), while others have modeled
isothermal flows with random forcing or
else in decaying regimes (Gammie \& Ostriker 1996; Padoan \& Nordlund
1999; Ostriker, Gammie \& Stone 1998; Mac Low et al.\ 1998). The
simulations in the latter group may themselves differ
in the scale at which the forcing is applied or
in the characteristic scales of the random initial conditions of the
turbulence. Given the large available parameter space, it becomes
necessary to constrain the parameters by comparing the simulation results with
suitable observational data. Efforts in this direction have recently
started to appear (Padoan, Jones \& Nordlund 1997; Heyer \& Schloerb
1997; Padoan et al.\ 1998; Padoan \& Nordlund 1999; Heyer \& Brunt
1999; Ballesteros-Paredes, Goodman \& \VS\ 1999; Rosolowsky et al.\
1999; Ossenkopf 1999).

The observations best suited for characterizing the turbulent
parameters of the ISM are densely sampled, high-resolution spectral
maps of molecular and atomic gas, such as the FCRAO CO Survey of the
Outer Galaxy (Heyer et al.\ 1998) and the Canadian Galactic Plane
Survey in HI (English et al.\ 1998). These provide information on the 
projected two-dimensional (on the plane of the sky) distribution of the 
gas, and
on the radial component of the velocity. However, because 
observational data are the result of a complex process which involves
integrating the 
radiation intensity along the line of sight (LOS) through a highly
inhomogeneous density field, convolving the result with instrumental
noise and response, and then ``selecting'' the intensity by its frequency 
to produce a spectrum, it is necessary to investigate the effects of this
process and the ``distortions'' it produces.

In this paper we take a first step towards understanding the
``translation'' of three-dimensional (3D) physical fields into
observational data by 
investigating the effects of integrating the density of 
a numerically-simulated field over portions of the LOS 
where the  LOS velocities lie in a certain interval,
in order to produce the analogue of spectral-line data from an
optically thin medium. Synthetic spectra have previously
been produced from simulations by Falgarone et al.\ (1994), Dubinsky,
Narayan \& Phillips (1995) and Padoan et
al.\ (1998), in order to compare the synthetic line profiles with
actual line spectra. However, in this paper we concentrate on the {\it
morphology} on the projection plane,
comparing the spatial distribution of the velocity channels with that 
of 2D slices of the original 3D density and LOS-velocity fields.

To this end, we use a 3D numerical simulation of compressible 
turbulence in the ISM, called ISM128,
analogous to the two-dimensional simulations in Paper I, and
including self-gravity, parameterized heating and cooling, star 
formation and rotation. In \S\ \ref{models}  we present
the equations, parameters and constants of the simulation. In \S\ 
\ref{3d_results} we present the 3D density and  
velocity fields in physical (real or configuration) space and discuss 
some of their physical properties. Next,
in \S\ {\ref{projection} we discuss the position-position-velocity
(PPV), or ``channel map'' representation of the data, noting that
important differences arise compared to the  
physical-space representation. Discussion of the effects of projection
is presented in \S\ \ref{discussion}. Finally, \S\ 
\ref{conclusions} contains the summary and conclusions.

\section{The Model} \label{models}

We solve the MHD equations in three dimensions as in the 
model for the ISM proposed in Paper I. We
consider a magnetized self-gravitating single fluid and include
model terms for cooling ($\Lambda$), diffuse background heating ($\Gamma_d$), 
local heating from stellar activity ($\Gamma_s$) and
shear. Specifically, we solve the following equations, at a resolution 
of $128^3$ grid points:
\begin{eqnarray}
\quad\frac{\partial\ln \rho}{\partial t}+\nabla\cdot(\ln\rho\bf u)
      +(1-\ln\rho)\nabla\cdot {\bf u}
=\mu(\nabla^2\ln\rho +
      (\nabla\ln\rho)^2) ,\label{mhd1}\\
%
\quad\frac{\partial{\bf u}}{\partial t}+{\bf u}\cdot\nabla{\bf u}=
     -\frac{\nabla p}{\rho} -{(\frac{J}{M})}^2\nabla\varphi
     -\nu _8\nabla ^8{\bf u} +\nu_2 (\nabla^2 \u + \frac{1}{3} \nabla
     \nabla \cdot \u)\nonumber \\ 
+\frac{1}{\rho}(\nabla\times{\bf B})\times{\bf B}
 -2\Omega\times{\bf u} \label{mhd2}\\
%
\quad\frac{\partial\ln e}{\partial t}
    +{\bf u}\cdot\nabla\ln e =
     -(\gamma -1)\nabla\cdot{\bf u}
+\frac{\kappa _T}{\rho}
      (\nabla ^2\ln e +(\nabla\ln e )^2)
+\frac{\Gamma _d+\Gamma _s+
      \rho \Lambda }{e},
      \label{mhd3}\\
%
\quad\frac{\partial{\bf B}}{\partial t}=
     \nabla\times ({\bf u\times B})-\nu _8\nabla ^8{\bf B}
  + \eta \nabla^2 {\bf B}\label{mhd4}\\
%
\quad\nabla ^2\varphi =\rho -1,\label{mhd5}\\
\quad P=(\gamma - 1)\rho e,\label{mhd6}
\end{eqnarray}
where $\rho$ is the density,  ${\bf u}$  the velocity, $P$ the thermal 
pressure, $\varphi$ the gravitational potential, $\Omega$  the angular
velocity due to galactic rotation,
${\bf B}$ the magnetic field and  $e$ the internal energy per unit mass.
In eq. (\ref{mhd3}) $\kappa _T$ is the thermal
diffusivity and  $\gamma$ the ratio $c_p/c_v$ of specific heats
at constant pressure and volume.
The temperature $T$ is related to $e$ by $e=c_vT$. 
Note that the continuity and the internal energy equations are solved in 
logarithmic variables. The mass diffusion term, in the right-hand side of the 
continuity equation (1) is added to smooth out density gradients, allowing
for greater Mach numbers than otherwise.
Momentum and magnetic field dissipation are included by using
a hyper-viscosity scheme with a $\nabla ^8$ operator, which 
confines viscous effects to the smallest scales and allows for much
smaller values of the second-order kinetic and magnetic diffusivities
$\nu_2$ and $\eta$ than would be required otherwise (see Paper I). 

The non-dimensional parameters which result from the normalization
are the Jeans number $J=\lambda _J/L_0=0.5$  and  the Mach number
$M=u_0/c_0=1$,
where $\lambda _J$ is the Jeans length and $c_0$ the adiabatic speed of
sound. Although the parameters for run ISM128 are the same as those of the
so-called run 28 in Paper I (except for those parameters which are
resolution-dependent, such as the viscosities and diffusivities), we
have adopted a new set of self-consistent physical units, more appropriate for
the 3D case, in which the spatial extention of the simulation box in
the vertical direction should not be so large as to bring in serious
stratification effects. We have thus re-scaled the units to a box size
of 300 pc and a code temperature unit $T_0=3000$ K. The chosen values
of the $J$ and $M$ parameters then imply a 
density unit of $n_0=3.3$ cm$^3$, a velocity unit of $u_0=6.41$ km s$^{-1}$
and a magnetic field unit of $B_0=5 \mu$G. In previous papers the domain 
size was 1 kpc, and the units were $T_0=10^4$ K, $n_0=1$ cm$^3$, $u_0=11.7$ 
km s$^{-1}$, with $B_0$ unchanged.

The equilibrium temperatures
$T_{\rm eq}$ as a function of density are obtained from balancing 
the heating and cooling functions. For example,
at $n=55$ cm$^{-3}$, $T_{\rm eq} = 100$ K, while at $n=0.1$ cm$^{-3}$,
$T_{\rm eq}=9,150$ K. Therefore, the new set of units still provides a
reasonably realistic account of the ISM for this range of densities and
temperatures.

In this model, the numerical box is centered at the solar Galactocentric
distance, with the directions $x$, $y$ and $z$ taken to correspond to 
the azimuthal, radial and vertical directions in the Galactic disk,
respectively. Note that, although the former coordinates refer to a 
Cartesian system and the latter to a cylindrical one, no conflict arises,
since locally, in the Solar neighborhood, the Galactic coordinates can be 
approximated by a Cartesian 
system. All projections are done along the z (vertical) direction, and thus 
we indistinctly refer to it as ``the line of sight'' (LOS) direction. 
Finally, the angular rotation velocity ${\bf \Omega}$ is taken equal to
that of the Sun, $|{\bf \Omega}|= (2\times 10^8 \ {\rm yr})^{-1}$.
 
Details concerning the choice of the model terms for heating and cooling
can be found in Paper I. The equations are solved using 
a pseudo-spectral method with periodic boundary conditions. 
The temporal scheme is a combination of a third order Runge-Kutta scheme
for the non-linear terms, and a Crank-Nicholson for the linear terms.
Initial conditions for all variables are Gaussian fluctuations with 
random phases, with no correlation between the various variables.

\section{Real-Space Representation (Density and Velocity
fields)} \label{3d_results}

In this section we discuss the density and $z$-velocity fields in three
{\it spatial} dimensions. For ease of presentation we have chosen to
show these 3D data by means of fixed-$z$ planes of the form $(x,y,z=z_0)$, 
rather
than attempting  to show the full 3D cubes. We refer to these planes 
as ``slices'' through the 3D spatial cubes, perpendicular to the $z$-axis.
Note that these slices correspond to a {\it single} value of $z$, contrary 
to the ``channel maps'' discussed in the following sections, which 
correspond to an {\it interval} of values of the $z$-velocity $u_z$.

Figure \ref{denvel}a shows 16 slices through the spatial 3D
density cube along the $z$-direction, with a separation of $\triangle z=7$ 
pixels between
slices. Whiter (darker) tones denote larger
(smaller) values of the field. Figure \ref{denvel}b shows the
$z$-component of the velocity field
($u_z$) on the same fixed-$z$ planes as in fig.\ \ref{denvel}a, with
black (white) meaning negative (positive) values of $u_z$.\footnote{The 
slices (i.e. constant-$z$ planes) through the $u_z$ cube (which exists in
three {\it spatial} dimensions) should not be confused with the 
velocity-channel maps discussed in the next section, which 
are the constant-$u_z$ planes of position-position-velocity (PPV) space. 
It is an unfortunate coincidence that PPV space is often refered to simply 
as velocity space, creating a potential source of confusion.}

The spatial structure in run
ISM128 exhibits large, interconnected high-density complexes, some of
them with abundant expanding shells (left side, central frames in
fig.\ \ref{denvel}a), and some others with not-so-large densities and
no star formation (lower parts, bottom frames in fig.\
\ref{denvel}a). The simulation gives the impression of a large degree
of global coherence. Structures along the z-direction can be identified 
as objects that persist over several frames. For example, the structure 
seen in the upper-right quadrant is actually a sheet, since it looks elongated 
(in the up-left to down-right direction), but it also extends from
$z=1$ to $z=50$. Then from $z=50$ to $z=71$, it is seen to contain an 
expanding shell (another small shell is seen at $z=8$).

\section{Position-Position-Velocity (Channel-Map) Representation}
\label{projection}

In order to study the effects of projection we have constructed
position-position-velocity (PPV) data ``cubes'' from the 3D density and
velocity fields (or
cubes). At each position $(x,y)$, the {\it velocity} axis contains a
density-weighted histogram of the $z$-velocity $u_z$. ``Channel maps'' are
then constructed as images of the column density with $z$-velocity between
$u_z$ and $u_z+\triangle u_z$ as a function of the remaining two spatial 
coordinates $x$ 
and $y$ (the $z$-direction is integrated out upon computing the column 
density). In brief, the channel
maps are slabs in the PPV cube of thickness $\triangle u_z$.

A precise definition can be given as follows. Let $\mu$ be
the measure defined by $\mu(E;x,y)=\int_E \rho(x,y,z)dz$  where
$E$ is a set of points in the LOS at a location $(x,y)$ in the
transverse plane. Consider the sets $\{ u_z<u\}=\{(x,y,z)\in
X\ |\ u_z(x,y,z)<u\}$, where $X$ is the whole integration domain. 
The LOS velocity cumulative distribution function
(``weighted by density'') is defined as $\mu(\{u_z<u\})= G(x,y,u)$.
The density-weighted LOS velocity histogram is thus
$dG/du$. The channel maps, or $(x,y)$ maps of column density
associated with points moving with an LOS velocity $u_z$ between $u$
and $u+\Delta u$ are obtained by just taking $(dG/du)\Delta u$.
We have the property that
the velocity-integrated (over $u_z$) maps 
are identical with the maps of the $z$ (spatial) projections of the
density field: $\int_{-\infty}^{\infty}(dG/du)du=\int_C
\rho(x,y,z)dz$, where $C$ is the whole LOS at $(x,y)$.
Figure\ \ref{PPV}a shows a map of the projected density field (i.e.
column density) along the $z$-direction.

In this paper we 
employ a relatively low velocity resolution of only 16 velocity channels.
However, since here we are not interested in distinguishing velocity
features, but rather in investigating the spatial structure seen in
the channel maps, such relatively low velocity resolution is
inconsequential. On the other hand, excessive velocity resolution
would cause poor sampling of the gas along the LOS, due to the low spatial
resolution of the simulation. The range of physical velocities 
is roughly from $-6$ to $+6$ km s$^{-1}$, and the channel width is therefore 
$\sim 0.75$ km s$^{-1}$. The 16 channel maps are shown in Fig.\ \ref{PPV}b.

Note that our choice of using 16 velocity channels and 16
slices through the 3D density and velocity fields is fortuitous,
though convenient, since it allows presenting them alongside each other.
However, we must emphasize that there is no direct relation between the
channel map data and the density or velocity data. While the former refer 
to fixed ($u_z,u_z+\triangle u_z$) intervals in PPV space, the density and velocity
slices refer to constant-$z$ planes in real (configuration) space. 
Furthermore, the spatial cubes have a size of $128 \times 128 \times 128$
spatial data points, while the PPV ``cubes'' (actually parallelepipeds) have 
a size of $128 \times 128$ 
spatial points $\times 16$ velocity channels.

An important issue to remark is that, due to the method used for their
construction, the channel maps may be interpreted in two
ways. The standard interpretation is as maps of the spatial distribution
of the column density in a
given LOS velocity interval. However, they can alternatively be interpreted as
{\it maps of the projected spatial distribution of a given velocity in
the LOS} weighted by column density. That is, in the former case one refers
to the (projected) spatial distribution of the {\it density} field, while in
the latter one refers to that of the $z$-velocity field. 
As we shall see in sec.\ \ref{discussion}, the latter interpretation appears 
to be at least as meaningful as the former.

\section{Effects of Projection} \label{discussion}

\subsection{Morphological similarity between the channel maps and the
original velocity field} \label{morpho_sim}

Comparing fig.\ \ref{PPV}b to figs.\ \ref{denvel}a and \ \ref{denvel}b, we
note that the morphology in the channel maps appears more similar to the
structure seen in the slices of the spatial $u_z$ cube than in those of 
the spatial density cube.
In order to quantify this effect, we have constructed histograms of the 
pixel-to-pixel (linear Pearson) correlation function between each
velocity channel and each density or velocity slice.
That is, for each image pair, we compute the correlation by considering
pairs of pixels (one pixel from each image) with the same
$(x,y)$ coordinates. This procedure thus estimates the pixel-to-pixel
morphological similarity between the two images. Unfortunately, this
procedure cannot be used for estimating the similarity between
images which do not represent the same object or physical system, since
in this case the spatial features in each image are not expected to
coincide.

Since the channel maps are LOS integrations while the density and
velocity images are slices through 3D spatial cubes, there are no
preferred pairs of images between which to compare. We have thus 
computed the correlation between all possible pairs of images, and
then plotted the histograms of the channel map-to-density slice and
channel map-to-$z$-velocity slice correlations. We refer to these as
the PPV-$\rho$ and PPV-$u_z$ correlation histograms, shown in
figure \ref{histograms}. We see that the histogram of the PPV-$u_z$
correlations contains many more large absolute values than the 
PPV-$\rho$ histogram. Note that both positive and negative PPV-$u_z$ 
correlations exist because $u_z$ can be either positive or negative 
while the PPV-$\rho$ correlations are positive-definite since both the 
``intensity'' in the channel maps and density are also positive definite. 
This seems to confirm the greater morphological similarity of the
$z$-velocity slices to the channel maps in a statistical sense. We
conclude that {\it the morphology seen in 
the channel maps is more representative of the projected spatial
distribution of $u_z$ than of the column density
distribution}.  However, it is important to emphasize that the above
conclusion does not imply that the channel maps give no information
about the projected density distribution, but only that the spatial
distribution of the LOS velocity seems to be more important than that 
of the density in the determination of the structure seen in the channel 
maps.

What is the origin of the above effect? To see this, consider an
incompressible flow. In this case, one
can still construct channel maps, in which all the structure  
is due to the spatial distribution of the velocity field, since
the density is a constant. Lazarian \& Pogosyan (1999) have independently
pointed this out, and numerical experiments to ``observe'' a
non-hydrodynamic, random velocity field with uniform density to
produce channel maps have been performed by Heyer \& Brunt (1999).
Conversely, consider now an inhomogeneous
flow but with constant velocity. In this case, one velocity channel will
contain the total projection of the density field. In the general case 
where neither the density nor the velocity are constants, the
structure seen in the channel maps is a complex mixture of the
structure in both the density and LOS-velocity fields. The 
unexpected result found here is that it is the morphology of the LOS
velocity field which tends to dominate the channel map structure, at
least in the flow regimes we have simulated.

\subsection{Small-scale structure}\label{small_scale}

A second result seen from figs.\ \ref{denvel}a and \ref{PPV}b is that 
the channel maps contain more small-scale structure than the original 
spatial density cube, exhibiting sharper structures, edges and
filaments, almost as if they had been made at significantly higher
resolution. 

In order to quantify this, we have computed the power spectra of the 
two dimensional (2D) density and $u_z$ slices, and of the channel maps 
(also 2D). The power spectrum is defined in two dimensions as

\centerline {$ E(k)= \frac{1}{2} \int |{\bf a_k}|^2 d\theta_k $}

\noindent where ${\bf a_k}$ is the Fourier transform of the 2D field, 
$\theta_k$ is the angular coordinate in the Fourier space, and {\bf k} 
is the wave vector with wave number {\it k}.  The spectra of the density 
and velocity slices labeled 57 in figs.\ \ref{denvel}a,b and the spectrum 
of the channel map labeled 9 in fig.\ \ref{PPV}b, are shown
in fig.\ \ref{spectra}, with respectively solid, dotted and dashed lines. 
For clarity we only have shown the spectrum of one representative frame 
from each cube, but the behaviour we describe is typical of all frames. 

Due to the low resolution of the simulation, only marginal power-laws 
are seen in the spectra of the density and velocity slices. These spectra 
curve down due to the mass-diffusion and viscous terms in the continuity (1) 
and momentum (2) equations, respectively. The density-slice spectrum enters 
this diffusive range at lower wavenumbers because of the lower order of the 
mass-diffusion term compared to that of the hyperviscous term. The channel 
map spectrum, on the other hand, is much closer to a power law, and, 
interestingly, does not curve down at large wavenumbers. This indicates 
that channel maps ``create'' their own small-scale structure.

The origin of this larger amount of small-scale power in the channel map
spectra might to be
due to the fact that density-weighted $u_z$ histograms for
nearby points in the projection plane
contain in general contributions from
points which may be far apart in configuration space. This effect is not
noticeable in the total projection because in this case all points
along an LOS are summed over. However, upon differentiation with respect to
velocity, the points included in any one LOS are only a small fraction 
of the total number available, and in general the segments contributing
to any given velocity bin may vary significantly
from one LOS to the next. This effect may cause the correlation length to 
be smaller in the channel maps than in the original density or LOS
velocity fields. To test for this mechanism, we have performed two
experiments: one in which we have replaced the actual LOS velocity
field of the simulation by an $(x,y)$-independent field of the form
$u_z(z)=z$, with $z$ in units of the integration box size. The second
experiment adds a sinusoidal dependence on the $(x,y)$ position of the 
LOS, setting $u_z(z)=z+2\pi \omega x + 2\pi \omega y$, with $\omega
=6$. We find that while the 
$(x,y)$-independent case shows a strong drop-off of the channel map spectra 
(not shown) at small scales, the second experiment restores their larger power
content. These experiments confirm that the origin of the small-scale excess 
power in the channel maps is the $(x,y)$-dependence of $u_z$, presumably 
because it causes the sampling of different sections along $z$ from
one LOS to the next. However, the reason why the channel maps do develop
power-law spectra, even when the original density and velocity fields
do not, remains uncertain to us.

A related effect is that localized structures in the channel maps do
not necessarily correspond to localized structures in physical space,
and viceversa. For example, the structure labeled A in the 11th
velocity channel in fig.\ \ref{PPV}b is seen to result from the
superposition of two density structures also labeled A in slices 29
and 71 of the 3D density field (fig.\ \ref{denvel}a).  Conversely,
expanding shells (due to the modeled star formation) are relatively
localized in physical space, but extended in PPV space, such as the
structure labeled B in slice 64 of the density field (seen since
slice 57), which extends from channels 8 through 12 in PPV
space. The ambiguity in the correspondence between real structures and
structures in PPV space has been previously pointed out by Issa,
MacLaren \& Wolfendale (1990) and Adler \& Roberts (1992).

One final cautionary note is in order. Throughout this paper we have 
projected and velocity-selected the density field without accounting 
for any possible thermal broadening, which would spread the contribution 
of any one fluid parcel in space over various velocity channels, probably 
smoothing the structure in the channel maps back again (M. Heyer, 1999, 
private communication).

\section{Summary and Discussion} \label{conclusions}

In this work we have presented a 3D simulation of compressible
turbulence in an
interstellar context (run ISM128). We have discussed the differences
between the representations 
of the data in physical and position-position-velocity (``PPV'' or
``channel-map'') spaces. Although the heating and cooling functions used
in run ISM128 make it non-isothermal, our results are expected to be
applicable to cold, nearly isothermal molecular gas observed in
optically thin lines as well, since the effects we discussed have a
geometrical origin. The main results are as follows:

1. The structure in the channel maps is morphologically closer 
to the structure in slices through the 3D spatial LOS-velocity ($u_z$) cube
than to that in slices through the 3D density cube. Indeed, the histogram
of the pixel-to-pixel correlations between all channel maps and 
all LOS-velocity slices extends to larger values
than the corresponding correlations between the channel maps and 
the density slices. This implies that channel maps are more representative
of the spatial distribution of the LOS velocity than
of the density field. Experiments testing the effect of varying the
density field on simulated channel maps, using uncorrelated density and
velocity fields, have been performed by Brunt \& Heyer (1999, private
communication), finding that changes in the channel map structure as
the imposed density field is varied are relatively minor.

2. The channel maps exhibit more small-scale structure than the
density slices, the former having power spectra that curve down
systematically (due to the low resolution), while the latter exhibit
shallow power laws. The larger amounts of small-scale power in the
channel maps may be due to the random
superposition of distant structures (in configuration space) with the
same velocity along the LOS, because neighboring positions in the
projected maps 
are likely to integrate over different segments of the LOS,
producing significant, though artificial, small-scale structure in the 
channel maps.

3. Another consequence of the previous result is that localization in
physical space does not necessarily correspond 
to localization in channel-map space, and viceversa, an ambiguity already 
pointed out by Issa et al. (1990) and Adler \& Roberts (1992).

The relationship between the density,
LOS-velocity and channel map spectra has been investigated
theoretically by Lazarian (1995) and Lazarian \& Pogosyan
(1999). In particular, the latter authors have found that, assuming
uncorrelated random density and velocity fields with prescribed spectra,
the power spectrum of the channel maps is dominated by the velocity
(rather than the density) fluctuations, provided that the density power 
spectrum is steep enough. This is in qualitative agreement with our 
result of greater morphological similarity between the $u_z$ slices and
the channel maps. However, quantitatively our $u_z$-slice and channel-map
spectra do not look very similar. In fact, at low wavenumbers, the slopes 
of the channel map spectra ($\sim -2.2$) are very similar to those of the 
{\it density} slices rather than to the spectra of the $u_z$ slices, which
are sistematically steeper. Furthermore, the artificial small-scale
generation in the channel maps does not seem to be accounted for
by Lazarian \& Pogosyan (1999), and it should be pointed out that these 
authors compare the power spectrum of the full 3D fields with that of
the intensity maps, while here we compare the channel map spectra to 
the spectra of the 2D density and LOS-velocity slices. 
Thus, the two procedures are comparing different quantities in principle.
A final difference is that our density and velocity fields, which satisfy
the fluid conservation equations, are highly correlated, while Lazarian 
\& Pogosyan (1999) assume uncorrelated fields. Further work is needed to
fully understand the differences and similarities between the two approaches.

A final comment is in order. The structural comparisons between
channel maps and slices through 3D density and velocity data we have
performed in this paper may seem somewhat odd, since it amounts to
comparing LOS-integrated data with non-integrated
information. However, such is the nature of the complex process of
making an observation, and the ability to infer 3D spatial information
from its projection in the plane of the sky requires understanding the 
relationship between the two.

\acknowledgements

We thank Mark Heyer and John Scalo for helpful comments and enlightening
discussions. The simulation was performed on the Cray C98 of IDRIS, France. 
We acknowledge partial financial support from grants UNAM/DGAPA IN119998
and UNAM/CRAY SC-008397 to E.V.-S., the ``CNRS Programme National:
Physique et Chimie du Milieu Interstellaire'' to T.P., and a
CONACYT/CNRS grant to E.V.-S.\ and T.\ P.

\normalsize{

} 

\vfill
\eject


\begin{figure}
\caption{a) Sixteen slices through the 3D spatial
density cube of run ISM128 (in logarithmic scale). The separation $\triangle z$
between slices is 7 pixels. The $z$-coordinate of each plane is given by the 
number at the bottom of each frame. b) Slices through the 3D LOS-velocity
cube through the same planes as in (a) above.}
\label{denvel}
\end{figure}

\begin{figure}
\caption{a) Total projection of the density field of run ISM128 
along the $z$-direction (i.e., an integrated column density map). Two
large complexes with strong events of star formation can be seen on the 
left hand side of the box. b) The sixteen channel maps of the 
position-position-velocity $128\times 128\times 16$ ``cube'' (actually,
a parallelepiped). The channel velocity width is $\triangle u_z=0.75\  {\rm
km \ s}^{-1}$.}
\label{PPV}
\end{figure}

\begin{figure}
\caption{Histograms of the pixel-to-pixel PPV-$\rho$ (solid line) 
and PPV-$u_z$ (dashed line) correlations for all image pairs. The 
PPV-$u_z$ histogram clearly contains a larger number of large
correlations (in absolute value) than the channel map-{\it vs.}-density histogram.}
\label{histograms}
\end{figure}

\begin{figure}
\caption{Power spectra of the 2D images corresponding to the density ({\it
dotted line}) and velocity ({\it dashed line}) slices labeled 57 in figs. 
\ref{denvel}a and b, and to the channel map ({\it solid line}) labeled 9 of 
fig.\ \ref{PPV}b. The spectrum of the channel map is seen to contain 
increasingly more small-scale power with increasing wavenumber than 
either the density or the LOS-velocity slice spectra.}
\label{spectra}
\end{figure}

\vfill
\eject

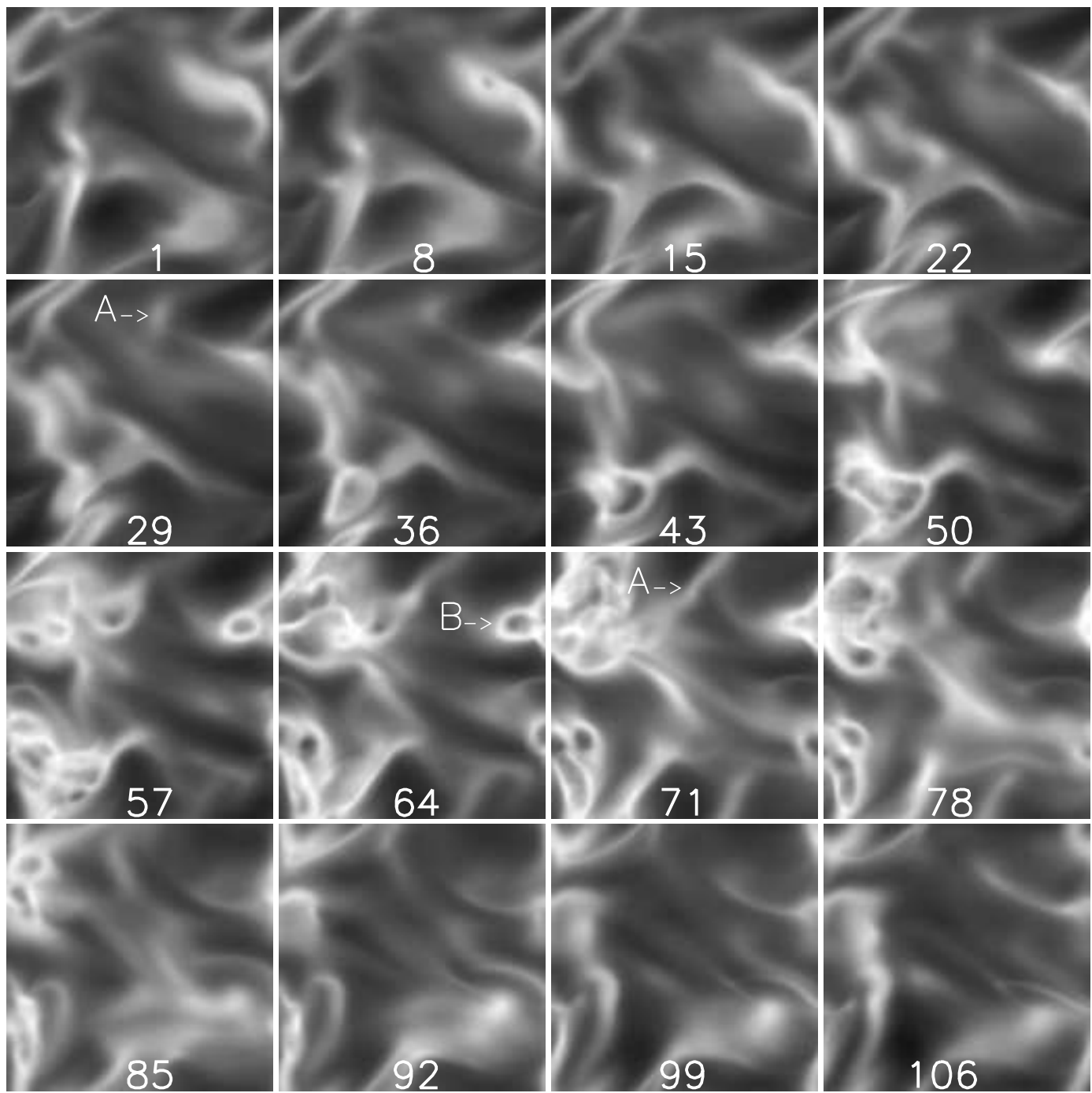
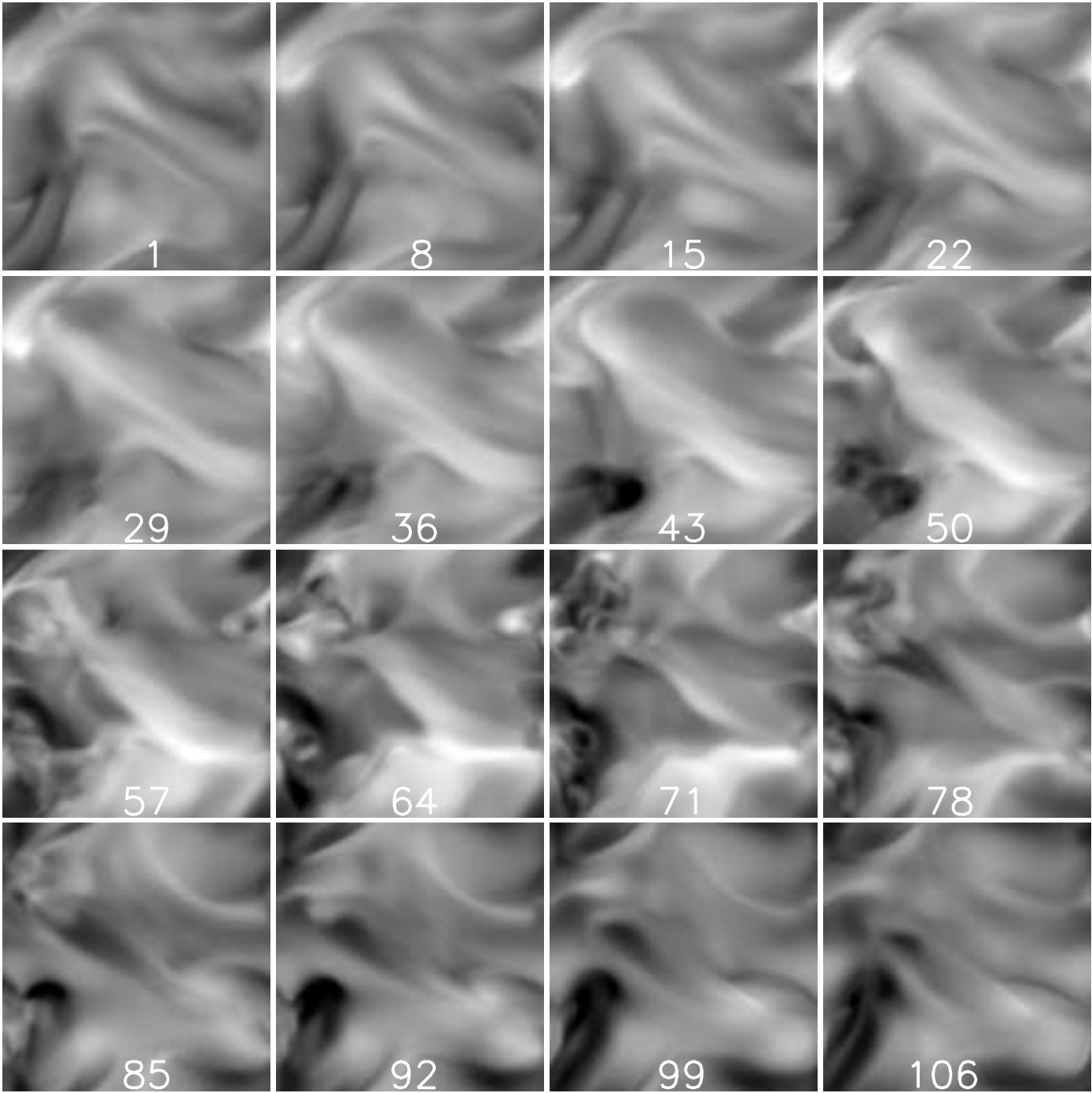
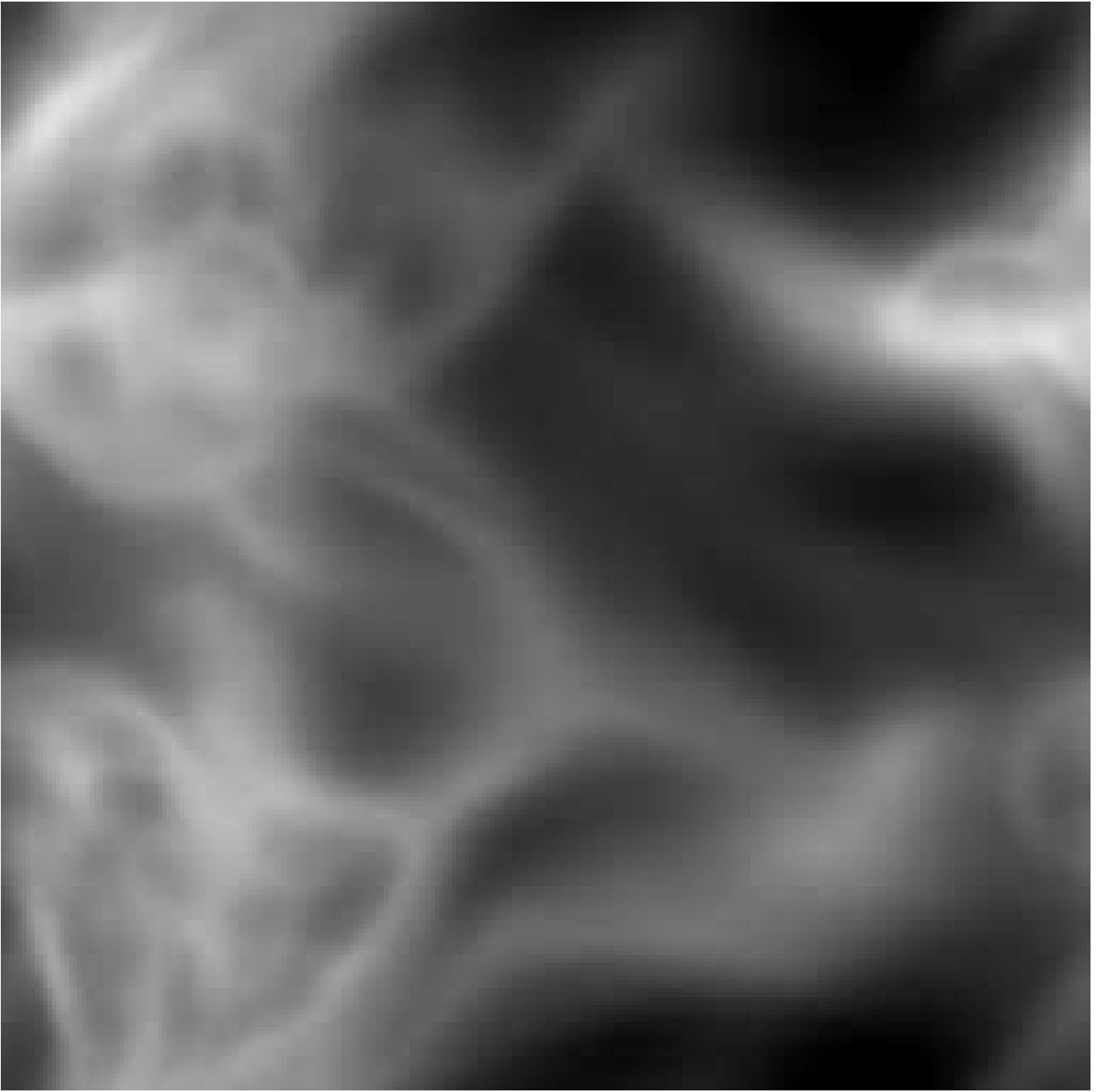
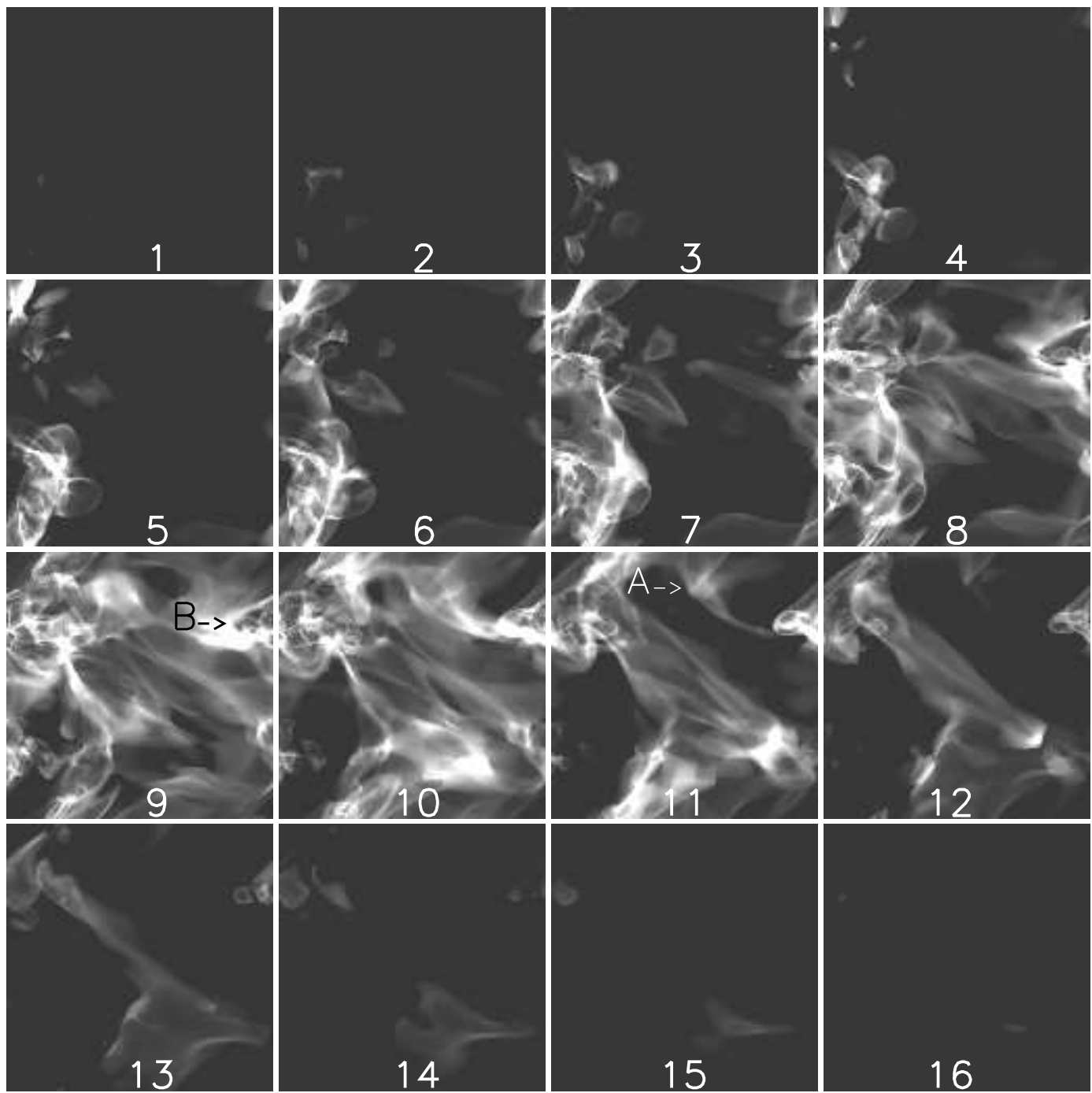
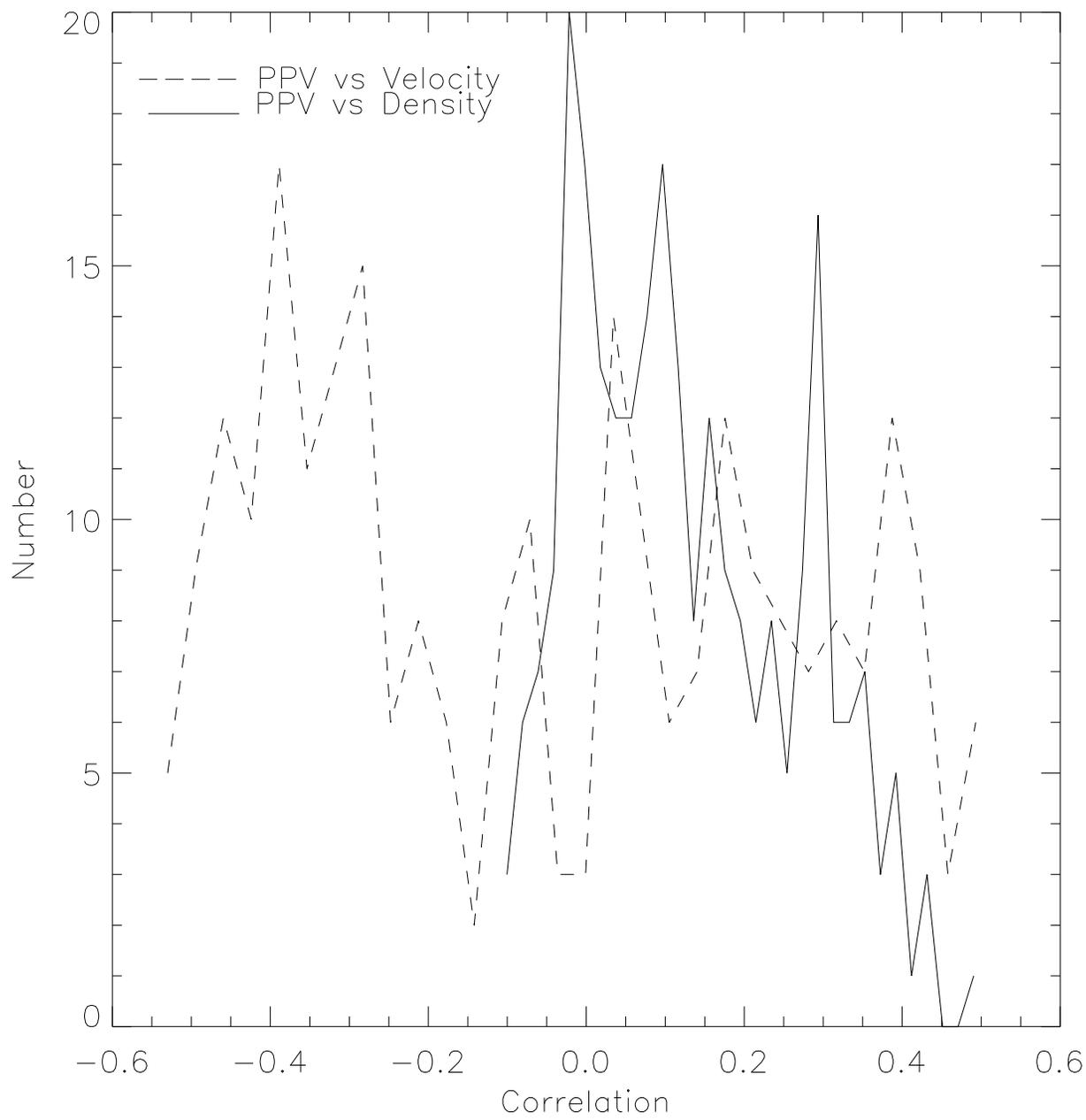
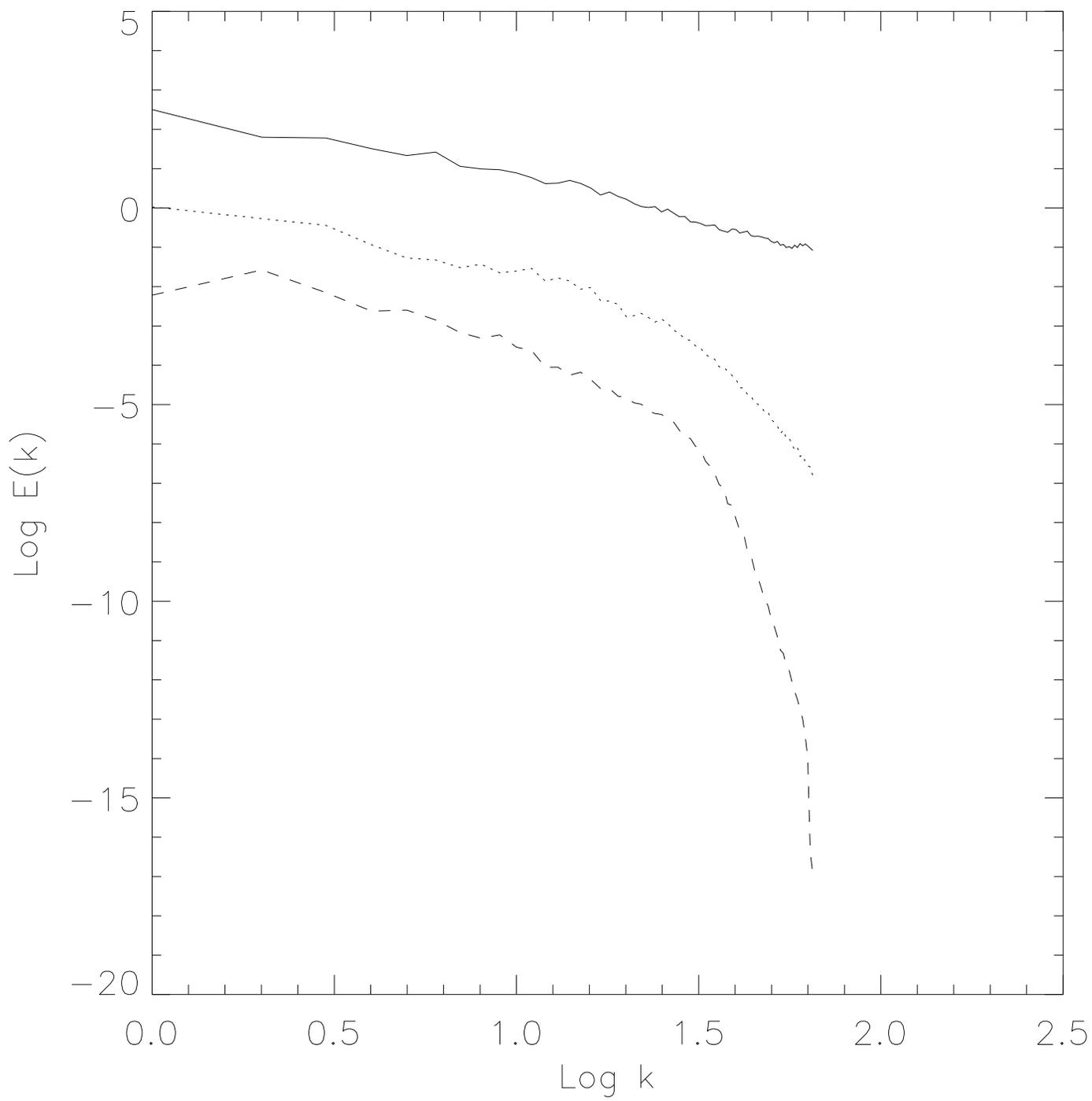
\end{document}